\documentclass[a4paper,twoside]{article}
\usepackage{units}
\usepackage{aas_macros}
\usepackage{dcolumn}
\usepackage{hyperref}
\usepackage{amssymb}
\newcommand{\affil}[1]{$^{\rm #1}$}
\usepackage[authoryear]{natbib}
\bibpunct{(}{)}{;}{a}{}{,}
\usepackage{graphicx}
\date{} 
\newcommand{\kms}{\mathrm{km\,s^{-1}}}
\newcommand{\rxjs}{RX\,J1605.3$+$3249}
\newcommand{\rxja}{RX\,J1856.5-3754}
\newcommand{\rxjn}{RX\,J0720.4-3125}
\let\orgautoref\autoref
\providecommand{\Autoref}
        {\def\equationautorefname{Equation}%
         \def\figureautorefname{Fig.}%
         \def\subfigureautorefname{Fig.}%
         \def\partautorefname{Part}%
         \def\chapterautorefname{Chapter}%
         \def\sectionautorefname{Section}%
         \def\subsectionautorefname{Section}%
         \def\subsubsectionautorefname{Section}%
         \def\Itemautorefname{Item}%
         \def\tableautorefname{Table}%
         \def\lstlistingautorefname{Listing}%
         \orgautoref}

\renewcommand{\autoref}
        {\def\equationautorefname{equation}%
         \def\figureautorefname{Fig.}%
         \def\subfigureautorefname{Fig.}%
         \def\partautorefname{part}%
         \def\chapterautorefname{chapter}%
         \def\sectionautorefname{section}%
         \def\subsectionautorefname{section}%
         \def\subsubsectionautorefname{section}%
         \def\appendixautorefname{appendix}%
         \def\Itemautorefname{item}%
         \def\tableautorefname{Table}%
         \def\lstlistingautorefname{Listing}%
         \orgautoref}
\providecommand{\autorefs}
        {\def\equationautorefname{equations}%
         \def\figureautorefname{Figs.}%
         \def\subfigureautorefname{Figs.}%
         \def\partautorefname{parts}%
         \def\chapterautorefname{chapters}%
         \def\sectionautorefname{sections}%
         \def\subsectionautorefname{sections}%
         \def\subsubsectionautorefname{sections}%
         \def\Itemautorefname{items}%
         \def\tableautorefname{Tables}%
         \def\lstlistingautorefname{Listings}%
         \orgautoref}
         
\makeatletter
\newcolumntype{d}[1]{>{\DC@{,}{{.}}{#1}}c<{\DC@end}}
\newcolumntype{o}[1]{>{\DC@{+}{\pm}{#1}}c<{\DC@end}}
\makeatother
\makeatletter
\newcolumntype{f}[1]{>{\DC@{p}{\ldots}{#1}}c<{\DC@end}}
\makeatother

\title{\large\bf\flushleft Neutron stars from young nearby associations -- the origin of \rxjs{}}
\author{\parbox{\textwidth}{\flushleft
\vspace{-0.5cm}
{\it Nina Tetzlaff\affil{A}, J\'anos G. Schmidt\affil{A}, Markus M. Hohle\affil{A}, and Ralph Neuh\"auser\affil{A}}\\
\vspace{0.4cm}
{\small \affil{A}\,Astrophysikalisches Institut und Universit\"ats-Sternwarte Jena, Schillerg\"asschen 2-3, 07745 Jena, Germany}}}
\begin{document}
\twocolumn[
\begin{changemargin}{.8cm}{.5cm}
\begin{minipage}{.9\textwidth}
\vspace{-1cm}
\maketitle
%
%
\small{{\bf Abstract}:

Many neutron stars (NSs) and runaway stars apparently come from the same regions on the sky. This suggests that they share the same birth places, namely associations and clusters of young massive stars. To identify NS birth places, we attempt to find NS-runaway pairs that could be former companions that were disrupted in a supernova (SN). The remains of recent ($<$few Myr) nearby ($<\unit[150]{pc}$) SNe should still be identifiable by observing the emission of rare radioisotopes such as $^{26}$Al and $^{60}$Fe that can also be used as additional indicators to confirm a possible SN event.
We investigated the origin of the isolated NS \rxjs{} and found that it was probably born $\approx\unit[100]{pc}$ far from Earth $\unit[0{.}45]{Myr}$ ago in the extended Corona-Australis or Octans associations, or in Sco OB4 $\approx\unit[1]{kpc}$ $\unit[3{.}5]{Myr}$ ago. A SN in Octans is supported by the identification of one to two possible former companions -- the runaway stars HIP 68228 and HIP 89394, as well as the appearance of a feature in the $\gamma$ ray emission from $^{26}$Al decay at the predicted SN place. Both, the progenitor masses estimated by comparison with theoretical $^{26}$Al yields as well as derived from the life time of the progenitor star, are found to be $\approx\unit[11]{M_\odot}$.}


\medskip
\medskip
\end{minipage}
\end{changemargin}
]
\small

\section{Introduction}

There are many young associations and clusters of massive stars in the solar vicinity that are potential birth places of neutron stars (NSs), hence supernova (SN) hosts. The NSs born in those SNe were ejected from their parent association or cluster shortly after formation due to a kick in an asymmetric SN explosion \citep[e.g.][]{1996PhRvL..76..352B,1996A&A...306..167J,2005ASPC..332..363J,2006ApJ...639.1007W,2009arXiv0906.2802K}. This scenario of NS kicks and ejection from its parent association or cluster is supported by the observation of large NS proper motions that indicate high space velocities \citep[e.g.][]{1994Natur.369..127L,1997MNRAS.289..592L,1997MNRAS.291..569H,1998ApJ...505..315C,2002ApJ...568..289A,2005MNRAS.360..974H}. 

\begin{figure}[h]
\begin{center}
\includegraphics[viewport = 70 0 500 840,width=0.25\textwidth,angle=-90]{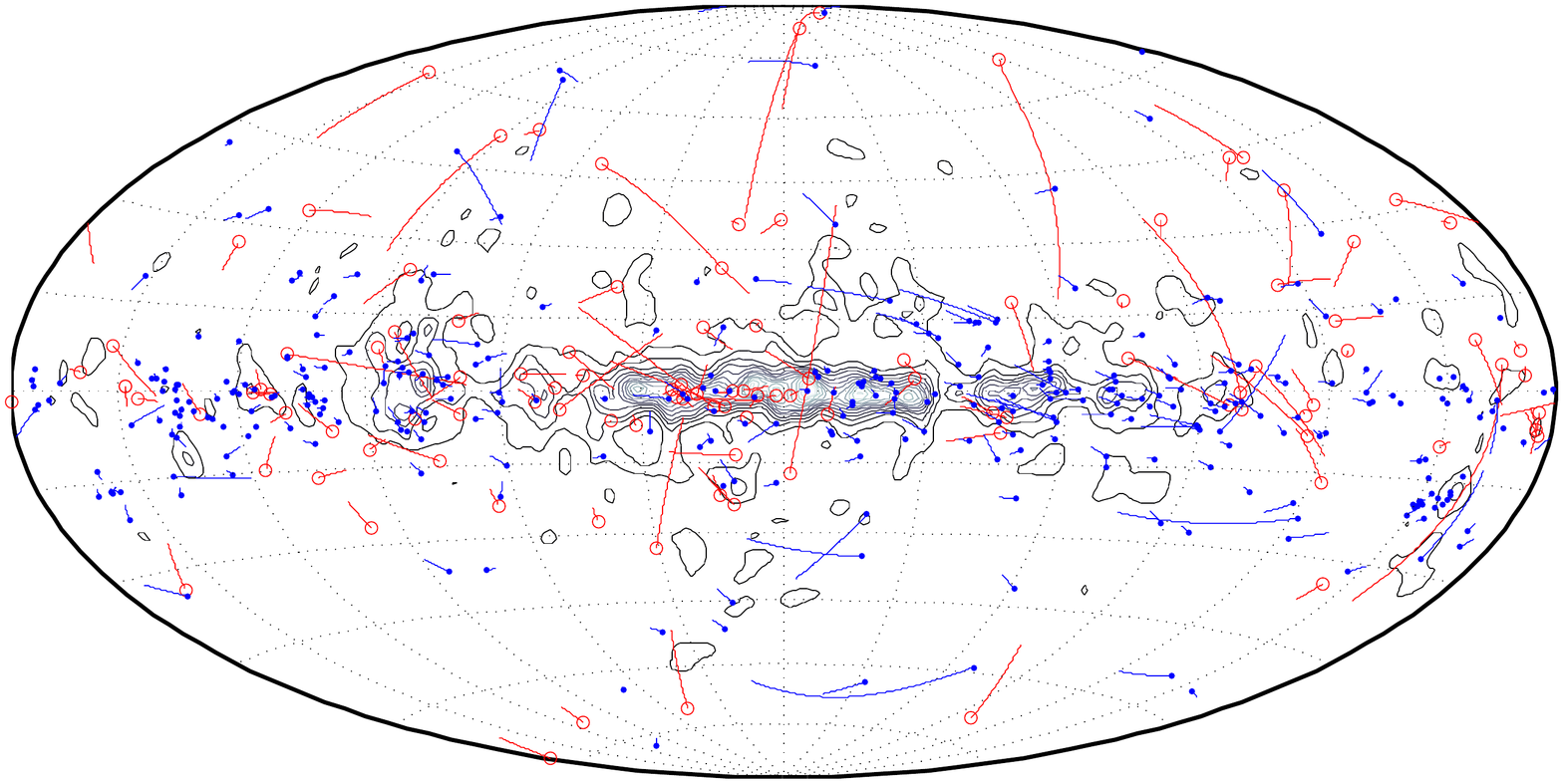}
\includegraphics[viewport = 70 0 500 840,width=0.25\textwidth,angle=-90]{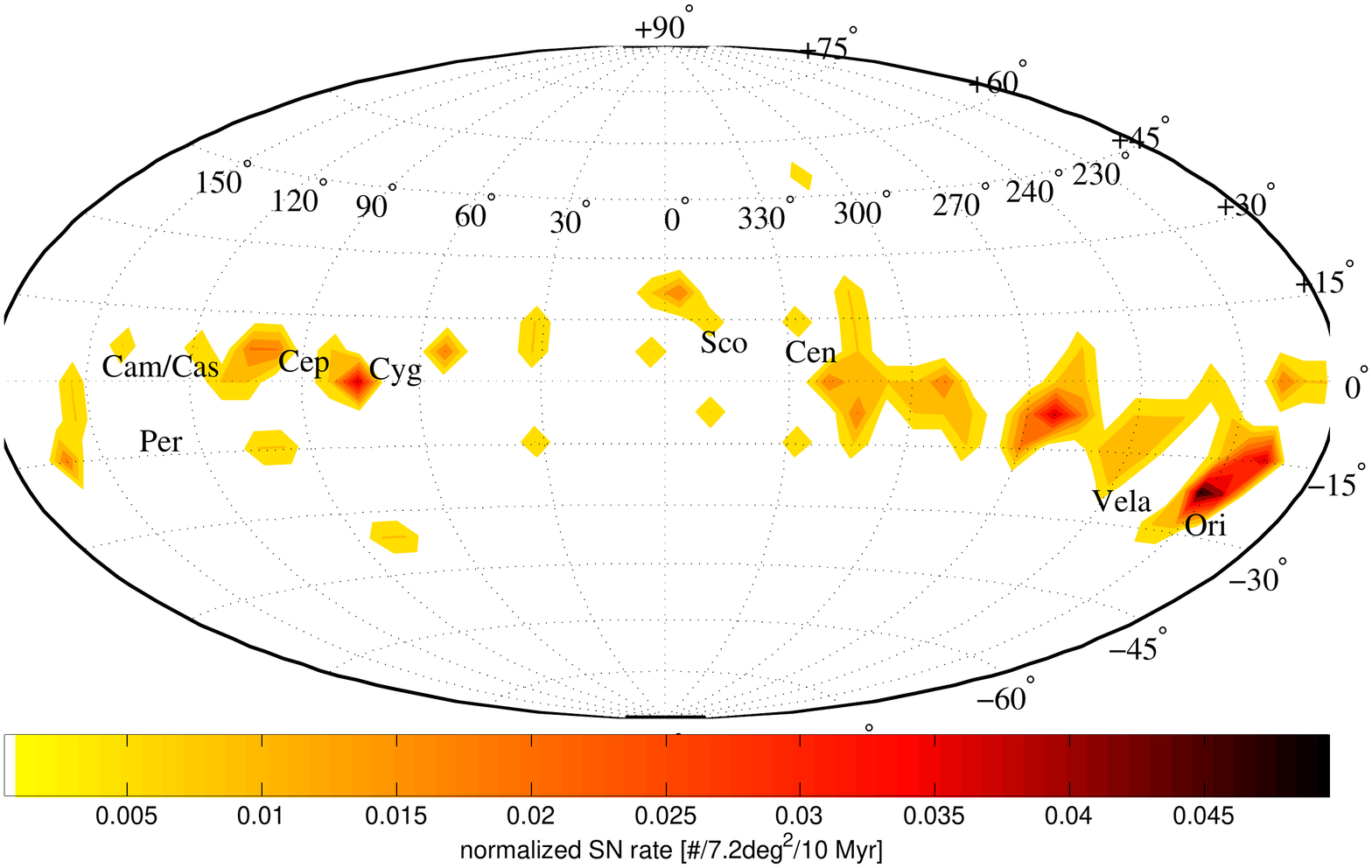}\vspace*{1em}
\caption{\label{fig:gammaplot} Top panel: Past flight paths ($\unit[1]{Myr}$ into the past) of young NSs (red, adopting zero radial velocity, kinematic data from ATNF pulsar database$^1$) and massive runaway stars (blue, { \citealt{2011MNRAS.410..190T}}). Symbols indicate their present position in Galactic coordinates. They are overplotted on the $^{26}$Al COMPTEL map from \citet{2010A&A...522A..51D} {(contours)}. { Most NSs and runaway stars seem to originate} from regions on the sky with enhanced $\gamma$ ray emission (e.g. Cygnus, Vela, Orion). Bottom panel: SN rate within $\unit[600]{pc}$ from the Sun (\citealt{Janos}; see also \citealt{2010AN....331..349H}), expected from current O and B type stars.}
\end{center}
\end{figure}

For a small number of NSs parent associations have been suggested \citep[e.g.][]{2001A&A...365...49H,2009MNRAS.400L..99T,2010MNRAS.402.2369T,2011arXiv1107.1673T,2008AstL...34..686B,2009AstL...35..396B}. Due to large uncertainties in the NS distances and the unknown radial velocities, the results are often not unique \citep{2010MNRAS.402.2369T,2011arXiv1107.1673T}. Therefore, further indicators are needed to decide on a particular birthplace. Such indicators may be the identification of a possible former companion that is now a so-called runaway star (\citealt{1961BAN....15..265B}, that should also show signs of the former binary evolution such as high helium abundance and high rotational velocity due to mass and momentum transfer from the primary as it filled its Roche lobe). Other indicators are sources of radioactive isotopes. Such isotopes are $^{26}$Al and $^{60}$Fe with half-lives of $\unit[0{.}72]{Myr}$ \citep[e.g.][]{1958ZNatA..13..847R,1984PhRvC..30..385T} and $\unit[2{.}62]{Myr}$ \citep{2009PhRvL.103g2502R}, respectively, i.e. much longer visible than a SN remnant ($\approx\unit[10^4]{yr}$). A cooling NS is visible for $\approx\unit[1]{Myr}$ \citep[see e.g. cooling curves in][]{2005MNRAS.363..555G,2009A&A...496..207P}, i.e. a similar time span. The comparison between maps of $\gamma$ ray emission, probable origins of runaway stars and NSs as well as the SN rate shows that there are regions on the sky where more SNe/ NSs are present than average (\autoref{fig:gammaplot}). Moreover, such regions contain young OB associations, e.g. Cygnus, Vela or Orion. Hence, it is plausible to search for NS and runaway origins within young associations and clusters. Providing small regions on the sky with enhanced SN/ NS number is also an important input for gravitational wave searches.

\footnotetext[1]{http://www.atnf.csiro.au/research/pulsar/psrcat/ \citep{2005AJ....129.1993M}}
\addtocounter{footnote}{1}
By tracing back in time young NSs, young runaway stars and young associations/clusters, we search for close encounters between the NS and a runaway star and/ or the NS and an association. To account for the errors on the observables as well as the for NS unknown radial velocity, we perform Monte Carlo simulations. 

If we find that a NS and a runaway star could have been at the same place and time inside an association/ cluster, it is well possible that the NS was born at that place at that time in a SN. From the flight time of the NS (its kinematic age), the association age and assuming contemporaneous star formation, we can also obtain the mass of the SN progenitor from its life time and evolutionary models. 
Converting the measured $\gamma$ ray flux due to $^{26}$Al decay into a mass of $^{26}$Al that was ejected in the SN can then test theoretical nucleosynthesis yields.

Here, after describing our method, we present results on \rxjs{}, a member of the so-called ''Magnificent Seven`` \citep[e.g.][]{2007Ap&SS.308..181H,2011heep.conf..345M,2011ApJ...736..117K}. \rxjs{} is a promising candidate for a nearby SN since its distance to the Sun is probably $\lesssim\unit[400]{pc}$ \citep{2007Ap&SS.308..217M}. Furthermore, for \rxjs{} the only way to give an estimate on its age (apart from comparison with cooling models) is to derive the age kinematically since no spin-down was measured so far and a possible $\unit[6{.}88]{s}$ spin period \citep{2007Ap&SS.308..181H} is currently not confirmed. Precise ages of NSs are also important to probe cooling models and constrain their composition, hence their equation-of-state.

\section{Procedure}\label{sec:proc}

We followed the same approach as already described in \citet{2010MNRAS.402.2369T} and \citet{2011arXiv1107.1673T} (see also \citealt{2001A&A...365...49H}). Therefore, we limit our description here to the most important facts.

We apply Monte Carlo simulations varying the parameters (parallax, proper motion and radial velocity) within their error intervals, to calculate the trajectories of the NSs and any association or cluster centre of a sample of 296 young associations and clusters within $\unit[3]{kpc}$ from the Sun into the past [see \citet{2011arXiv1107.1673T} for a description of the association sample, for associations listed in \citet{2009MNRAS.400..518M}, we took their parameters; here we add the Pleiades B1 moving group \citep{1999A&A...350..434A} since it was proposed to have hosted recent nearby SNe \citep{2002A&A...390..299B}]. 
At every time step ($10^4$ up to $\unit[5\times10^6]{yr}$ in steps of $\unit[10^4]{yr}$) the separation $\Delta$ between the association/cluster centre and the NS is calculated. We then find the minimum separation $\Delta_{min}\left(\tau\right)$ and the associated time $\tau$ in the past. For the radial velocity of the NS, we assume a reasonable probability distribution derived from the pulsar space velocity distribution by \citet{2005MNRAS.360..974H}. 

The distribution of separations $\Delta_{min}$ is supposed to obey a distribution of absolute differences of two 3D Gaussians with $\left(m_1,s_1\right)$ and $\left(m_2,s_2\right)$ being their expectation values and standard deviations (if the positional errors were Gaussian distributed), see \citet{2001A&A...365...49H},
{\scriptsize
\begin{equation}
W_{3D}\left(\Delta\right) = \frac{\Delta}{\sqrt{2\pi}s m}\left\{exp\left[-\frac{1}{2}\frac{\left(\Delta-m\right)^2}{s^2}\right]-exp\left[-\frac{1}{2}\frac{\left(\Delta+m\right)^2}{s^2}\right]\right\},
\label{eq:3DGauss_diff}
\end{equation}
}
where $\Delta$ denotes the 3D separation between two objects (here, NS and association centre or NS and runaway star; $\Delta=\Delta_{min}$), $m=\left|m_1-m_2\right|$ and $s^2=s_1^2+s_2^2$; $\pi=3.1459\ldots$.\\
For investigating encounters with runaway stars, we calculate $\Delta_{min}$ between the NS and the runaway star. If the two stars once were at the same place, i.e. $m \rightarrow 0$, \autoref{eq:3DGauss_diff} becomes 
\begin{equation}
W_{3D,m\to0}\left(\Delta\right) = \frac{2\Delta^2}{\sqrt{2\pi}s^3}\exp\left[-\frac{\Delta^2}{2s^2}\right].
\label{eq:3DGaussmunull_diff}
\end{equation}
Note that even in this case, there is zero probability of finding a value of $\Delta_{min}=0$ in the Monte Carlo simulation. However, near-zero values should be found after a sufficient number of runs (see \autoref{appsec:numMCruns}). The peak of the distribution is shifted towards larger $\Delta_{min}$ depending upon $s$. As the actual (observed) case is different from this simple model (no 3D Gaussian distributed positions, due to e.g. the Gaussian distributed parallax that goes into the position reciprocally, complicated radial velocity distribution, etc.), we will fit the theoretical formulae only to the first part of the $\Delta_{min}$ distribution (up to the peak plus a few more bins) such that the slope and peak can be explained (in the following, we use the term ``adapt'' instead of ``fit'' because \autorefs{eq:3DGauss_diff} and \ref{eq:3DGaussmunull_diff} are only used to explain the first part of the $\Delta_{min}$ distribution, hence it is not a real fit). The parameter $m$ then gives the positional difference between the two objects. Note that the uncertainties on the separation are dominated by the kinematic uncertainties of the NS that are typically of the order of a few hundred $\kms$ (because of the assumed radial velocity distribution). As a consequence, the distribution of separations $\Delta_{min}$ shows typically a large tail for larger separations. However, the first part of the $\Delta_{min}$ distributions (slope and peak) can still be explained well with \autoref{eq:3DGauss_diff} since the kinematic dispersions for only those runs are much smaller, typically a few tens of $\kms$ for the NS, i.e. a few tens of pc after $\unit[1]{Myr}$.\\

To associate the NS/runaway star encounter position with an association/cluster, the trajectory of the association/cluster is calculated simultaneously. Runaway star data are taken from \citet{2011MNRAS.410..190T} (and references therein, mainly \citealt{2007AA...474..653V}) for 2547 runaway stars (1705 with full 3D kinematics).\\

In general, we first perform $10^4$ Monte Carlo runs for each NS/association pair and $10^3$ Monte Carlo runs for each NS/runaway star pair (for the latter less runs are still sufficient due to the smaller errors on the runaway star kinematics compared to the dispersion of the association velocities) to find those associations and runaway stars that potentially crossed the past path of the NS, i.e. those for which the smallest $\Delta_{min}$ value found in the calculations is less than three times the association radius (for NS/association pairs) or less than $\unit[10]{pc}$ (for NS/ runaway star pairs), respectively (see \autoref{appsec:numMCruns}). Those associations and runaway stars that fulfilled those conditions are then selected for a more detailed investigation (one to three million Monte Carlo runs). The outcome of these simulations is then discussed in detail. Regarding associations, we search for those for which the NS could have been within the association boundaries in the past while for runaway stars, we are looking for those runaway stars for which the NS and the runaway star might have been at the same place in the past, hence the distribution (slope) should obey \autoref{eq:3DGaussmunull_diff}. After three million runs the smallest $\Delta_{min}$ value found is expected to be smaller than one parsec (see \autoref{appsec:numMCruns}). If this criterion is satisfied, we adapt \autorefs{eq:3DGauss_diff} and \ref{eq:3DGaussmunull_diff} to the first part of the ``observed'' $\Delta_{min}$ distribution to explain its slope and peak. If we find that a distribution with $m=0$ (\autoref{eq:3DGaussmunull_diff}) can satisfactorily explain the slope and peak of the $\Delta_{min}$ histogram, we consider the runaway star as former companion candidate. \\

This procedure was already successfully applied by \citet{2001A&A...365...49H}, \citet{2008AstL...34..686B,2009AstL...35..396B} and us \citep{2009MNRAS.400L..99T,2010MNRAS.402.2369T,2011arXiv1107.1673T} and was also applied to several (artificial) test cases \citep{2009DiplA...Nina}.

\section{\rxjs{}}

We adopt the following parameters for \rxjs{} (right ascension $\alpha$, declination $\delta$, distance $d$, proper motion $\mu_{\alpha}^* = \mu_\alpha\cos\delta$, $\mu_\delta$):

\begin{equation}
	\begin{array}{l c l}	
	\alpha &=&  16^\mathrm{h}05^\mathrm{m}18^\mathrm{s}\hspace{-0.8ex}.5,\ \delta\ =\ +32^\circ49\mathrm{'}17\mathrm{''}\hspace{-0.8ex}.4\\
					& &	\mbox{(\citealt{2003ApJ...588L..33K})},\\
	d &=& \unit[350\pm50]{pc}\ \mbox{(\citealt{2007Ap&SS.308..171P}\footnotemark)},\\
	\mu_{\alpha}^* &=& \unit[-43{.}7\pm1{.}7]{mas/yr},\ \mbox{(\citealt{2006A&A...457..619Z})}\\
	\mu_{\delta} &=& \unit[148{.}7\pm2{.}6]{mas/yr}\ \mbox{(\citealt{2006A&A...457..619Z})}.
	\end{array}\label{eq:1605input}
\end{equation}
\footnotetext{\citealt{2007Ap&SS.308..171P} used different models for the hydrogen column density and derived values of $\unit[390]{pc}$ and $\unit[325]{pc}$. \citet{2007Ap&SS.308..217M} give an upper limit of $\unit[410]{pc}$. According to \citealt{2007Ap&SS.308..171P}, the different models give consistent, hence very reliable results up to $\approx\unit[270]{pc}$ (their values are also in good agreement with the parallactic distances for the two NSs \rxja{} and \rxjn{}). Hence, our adopted distance for \rxjs{} of $\unit[350\pm50]{pc}$ is reasonable.} 
 
First, we perform $10^4$ Monte Carlo runs to find close encounters between \rxjs{} and any association/cluster in the past five million years\footnote{The effective temperature of \rxjs{} of $kT=\unit[96]{eV}$ \citep{1999A&A...351..177M,2004ApJ...608..432V} imply a cooling age of $\unit[10^5-10^6]{yr}$ \citep[see e.g. cooling curves by][]{2005MNRAS.363..555G,2009A&A...496..207P}, hence older ages can be excluded.}. We select those associations/clusters for which the smallest separation $\Delta_{min}$ found was less than three times the association/cluster radius $R_{ass}$, 18 in total. For those 18 associations/clusters, we carry out another one million Monte Carlo runs. For 10 of them we find close encounters consistent with the association/cluster boundaries. They are listed in \autoref{tab:assoc1605}. We adapt \autoref{eq:3DGauss_diff} to the first bins of each $\Delta_{min}$ distribution to obtain the distance of the SN to the association centre. Comparing the radii of each association with this putative separation of the SN from the association centre ($m\pm s$), three associations/clusters are found to be potential birth places of \rxjs{}, i.e. the position of the SN is consistent with the association boundaries within its standard deviation: the extended Corona-Australis association (Ext. R CrA), Scorpius OB4 and the Octans association. In \autoref{tab:assoc1605_2} we give the position of the SN and the properties \rxjs{} would currently have if it was born in the respective association. In the last Column we give an estimate of the mass of the progenitor star derived from the life-time of the progenitor star [association age (Ext. R CrA: $\unit[10-15]{Myr}$ \citealt{2000A&AS..146..323N}; Octans: $\approx\unit[20]{Myr}$, \citealt{2008hsf2.book..757T}; Sco OB4: $\unit[7]{Myr}$ \citealt{2005A&A...438.1163K}) minus NS age ($\tau$)] and evolutionary models from \citet{1980FCPh....5..287T}, \citet{1989A&A...210..155M} and \citet{1997PhDT........31K}. We regard these three associations as likely birthplaces of \rxjs{}. The derived uncertainties in the parameters (\autoref{tab:assoc1605_2}) are dominated by the uncertainties in the NS radial velocity and distance. They are included in the formal errors (68\%). For more distant associations ($>$few hundred pc) such as Sco OB4, the results are also influenced by the uncertainties in the association distances. These mostly affect the derived position of the SN within the association. Note that although the proposed progenitor mass of $\unit[42-89]{M_\odot}$ is rather high (because the lifetime of the progenitor star is short assuming it formed in Sco OB4 at the same time the association formed) and black holes are expected to form above $\approx\unit[25]{M_\odot}$ \citep{2003ApJ...591..288H} rather than NSs, for masses higher than $\approx\unit[50-80]{M_\odot}$ in binary systems, NSs are expected to form \citep{2008ApJ...685..400B}.\\
\begin{table}
\centering
\caption{Properties of potential close encounters between \rxjs{} and the centres of 10 associations/clusters.}\label{tab:assoc1605}
\scriptsize
\begin{tabular}{l d{0.0}  >{$}c<{$} c}
\hline
Assoc./	&	\multicolumn{1}{c}{$R_{ass}$}			&	\multicolumn{1}{c}{$\left(m, s\right)$}	&	$\tau$\\
cluster  &	\multicolumn{1}{c}{[pc]}			&	[pc]	&	[Myr]\\\hline
Tuc-Hor		&	50																											&	\left(130{.}4, 65{.}5\right)						&	$0{.}63^{+0{.}90}_{-0{.}11}$\\
$\beta$ Pic-Cap&			57																						&	\left(135{.}2, 59{.}0\right)						&	$0{.}60^{+0{.}07}_{-0{.}12}$\\
ext. R CrA& 31																										&	\left(33{.}4, 2{.}3\right)							&	$0{.}42^{+0{.}07}_{-0{.}04}$\\
AB Dor		&	43																									&	\left(155{.}0, 51{.}1\right)						&	$0{.}55^{+0{.}06}_{-0{.}10}$\\
Sco OB4$^\mathrm{a}$	&	30																		&	\left(42{.}7, 18{.}4\right)							&	$3{.}4^{+0{.}3}_{-0{.}6}$		\\
Columba		&	85																								&	\multicolumn{1}{c}{$m>200$}& $\approx0{.}6$							\\
Carina		&	58																									& \left(129{.}2, 60{.}0\right)						&	$0{.}68^{+0{.}11}_{-0{.}12}$\\
Octans		&	111																									&	\left(103{.}5, 25{.}7\right)						&	$0{.}53^{+0{.}09}_{-0{.}08}$\\
Argus	 		&	74																								& \left(148{.}7, 52{.}6\right)						&	$0{.}55^{+0{.}07}_{-0{.}09}$\\
Pleiades B1& 54																									&	\left(63{.}0, 4{.}1\right)							&	$0{.}67^{+0{.}13}_{-0{.}15}$\\
\hline
\multicolumn{4}{p{0.45\textwidth}}{$\left(m, s\right)$ -- expectation value and standard deviation of the encounter separation inferred from \autoref{eq:3DGauss_diff}, putative SN distance from association centre ($m\pm s$); $\tau$ -- time of the encounter in the past.\newline
$^\mathrm{a}$Using the \citet{2005MNRAS.360..974H} distribution for NS space velocities in the case of Sco OB4 we found that the absolute radial velocity $v_r$ of \rxjs{} must be very small if Sco OB4 was the birth association of \rxjs{}. To achieve better statistics and a clear peak in the $\Delta_{min}$ distribution to be able to adapt the theoretical curves, we repeated the calculations assuming a uniform $v_r$ distribution. The results for the latter are given here.}
\end{tabular}
\end{table}
\begin{table*}
\centering
\caption{Potential parent associations of \rxjs{}.}
\label{tab:assoc1605_2}
\setlength\extrarowheight{3pt}
\tiny
\begin{tabular}{c >{$}c<{$} >{$}r<{$} >{$}r<{$} o{4.2} o{4.2} >{$}r<{$} >{$}r<{$} >{$}c<{$} >{$}r<{$} >{$}r<{$} >{$}c<{$}}
\hline
Assoc.	&	\left(m, s\right) & \multicolumn{1}{c}{$\tau$} &	\multicolumn{1}{c}{$v_r$}			& \multicolumn{1}{c}{$\mu_{\alpha}^*$} & \multicolumn{1}{c}{$\mu_{\delta}$} & \multicolumn{1}{c}{$v_{sp}$} &	\multicolumn{1}{c}{dist}						&	d_{\odot} 	& \multicolumn{1}{c}{$\alpha$}	& \multicolumn{1}{c}{$\delta$} & M_{prog}\\ 
		& [pc] & \multicolumn{1}{c}{[Myr]}	& \multicolumn{1}{c}{[$\kms$]} & \multicolumn{1}{c}{[mas/yr]} & \multicolumn{1}{c}{[mas/yr]} & \multicolumn{1}{c}{[$\kms$]} & \multicolumn{1}{c}{[pc]} & \multicolumn{1}{c}{[pc]} & \multicolumn{1}{c}{[$^\circ$]} & \multicolumn{1}{c}{[$^\circ$]} & \mathrm{[M_\odot]}\\\hline
Ext. R CrA					& \left(33{.}4, 2{.}3\right)	& 0{.}42^{+0{.}07}_{-0{.}04}	& 577^{+123}_{-76} & -43{.}9+1{.}7 & 148{.}4+2{.}6 & 612^{+128}_{-72} & 303^{+30}_{-33}	& 104^{+7}_{-12}			& 261{.}5^{+1{.}2}_{-0{.}9}	& -44{.}6^{+2{.}3}_{-1{.}2} & 12-18\\
Sco OB4$^\mathrm{a}$			& \left(42{.}7, 18{.}4\right)& 3{.}4^{+0{.}3}_{-0{.}6}	& -21^{+19}_{-10}& -42{.}3+1{.}3	& 149{.}8+2{.}5	& 266^{+38}_{-33} & 385^{+50}_{-62}& 983^{+30}_{-12}			& 257{.}8^{+0{.}7}_{-0{.}5}	& -34{.}3^{+0{.}9}_{-0{.}8} & 42-89\\
Octans	& \left(103{.}5, 25{.}7\right)& 0{.}53^{+0{.}09}_{-0{.}08}		&  548^{+159}_{-34}& -43{.}6+1{.}7	& 148{.}6+2{.}6	& 664^{+87}_{-107} & 370^{+44}_{-35}& 140^{+6}_{-19}			& 285{.}0^{+17{.}3}_{-4{.}8}	& -78\ldots-68 & 10-11\\
\hline
\multicolumn{12}{p{\textwidth}}{$\left(m, s\right)$ -- expectation value and standard deviation of the encounter separation inferred from \autoref{eq:3DGauss_diff}, putative SN distance from association centre ($m\pm s$); $\tau$ -- encounter time. Predicted present NS parameters: radial velocity $v_r$, proper motion $\mu_\alpha^*$ and $\mu_\delta$, current distance $dist$; $v_{sp}$ -- predicted space velocity $v_{sp}$. Predicted SN position: SN distance $d_\odot$ (at the time of the SN), J2000.0 coordinates (as seen from Earth at present). Error bars denote 68 per cent confidence (cf. Appendix B in \citealt{2010MNRAS.402.2369T}). $M_{prog}$ -- progenitor mass\newline
$^\mathrm{a}$The results for Sco OB4 were obtained using a uniform radial velocity distribution for the NS instead of the one derived from the \citet{2005MNRAS.360..974H} distribution, see also comment in \autoref{tab:assoc1605}.}
\end{tabular}
\end{table*}
%


After $10^3$ Monte Carlo runs 40 runaway stars with full kinematics from \citet{2011MNRAS.410..190T} show a smallest separation $\Delta_{min}$ between \rxjs{} and the runaway star of less than $\unit[10]{pc}$. For them, another three million runs are performed. We select those stars for which separations smaller than one parsec are found after three million runs, 28 in total. To further reduce the number of potential former companion candidates, we follow the idea of \citet{2010AstL...36..116C}. They used a reference probability as significance indicator that two objects approach to small distances taking into account the spatial density distribution of the objects (we refer to details on this method to \citealt{2010AstL...36..116C}). \citet{2010AstL...36..116C} investigated pairs of pulsars whereas we investigate NS-runaway star pairs. Therefore, beside the spatial density distribution of NS, also the density distribution of runaway stars is needed. As our region of interest lies in the Solar neighbourhood, the pulsar distribution used by \citet{2010AstL...36..116C} is not sufficient for our purposes as it does not account for an increased SN rate in the Solar neighbourhood and includes all NSs (from young to very old). For that reason, we use the spatial distributions of young NSs and runaway stars derived by \citet{2012DokA...Nina}. They performed a population synthesis where NSs and runaway stars are ejected from their parent associations and clusters. As parent associations and clusters, an observed sample of associations and clusters was taken and member stars distributed according to the initial mass function from \citet{2005ASSL..327..175K}. SN times were then obtained from the age of the associations and the progenitor lifetime (for details we refer to \citealt{2012DokA...Nina}). From the present spatial positions of simulated NSs and runaway stars, their spatial distributions were derived.

From the Monte Carlo simulations we obtain the region $\left(X+\Delta X,Y+\Delta Y,Z+\Delta Z\right)$ in which the NS and runaway star can reach separations $\Delta_{min}$ no larger than $\rho$. If $\rho\ll\Delta\left(X,Y,Z\right)$, the number of objects in the volume $\rho^3$ obeys a Poisson distribution. Thus, the spatial region $\left(X+\Delta X,Y+\Delta Y,Z+\Delta Z\right)$ is divided into $N$ cells ($k=1\ldots k$) with edge lengths $\rho$ (see \citealt{2010AstL...36..116C}). Using the density distribution of runaway stars and NSs derived earlier, the number of stars ($\eta_{NS}$, $\eta_{run}$) in each cell is calculated and an average number in a cell of size $\rho$ is obtained. The probability of detecting $N$ objects of population $i$ (NSs or runaway stars) in a volume $\rho^3$ is given by
\begin{equation}
\hat{P}_0\left(\rho\right)  =  \sum\limits_{n_{i}=0}^{N}\frac{\eta_{i}^{n_i}}{n_i!}e^{-\eta_{i}}.
\end{equation}
The probability of detecting at least one NS and one runaway star in a volume with size $\rho^3$ is then given by
\begin{equation}
P_0\left(\rho\right) = \left(1-e^{-\eta_{NS}}\right)\left(1-e^{-\eta_{run}}\right).
\end{equation}
The probability of occurrence of separations $\Delta_{min}\leq\rho$ in the Monte Carlo simulations is $P\left(\rho\right)$. An upper limit $P_{0,upp}$ of $P_0$ can be derived from the Poissonian error on the calculated number of stars in each cell. If $P_{0,upp}<P$ for small $\rho$ (smaller than a few pc), this is an indicator that the association of \rxjs{} with a particular runaway star might be real.

Using this as a criterion, none of the associations between \rxjs{} and the former 28 companion candidates is found to be highly significant. However, for those cases, where the encounters are found within an association, the reference distribution for NSs and runaway stars \citep{2012DokA...Nina} predicts a relatively high number of stars, as expected. Hence, those encounters are expected to be found less significant. Therefore, we still regard runaway stars for which encounters with \rxjs{} are found within one of the potential parent associations of \rxjs{} listed in \autoref{tab:assoc1605_2}, namely the Ext. R CrA, Sco OB4 or the Octans association, as good candidates.\footnote{Note, that there might have been associations or clusters in the past that are already dispersed. However, since \rxjs{} is very young, its parent association or cluster should still exist as the dispersion time scale of associations and clusters is much larger (tens to hundreds of Myr).} Here, for the Ext. R CrA and Octans associations, we do not only regard the nominal radii (maximum extensions) of the associations but their extensions in different directions in space taking into account the motion of the association (for extensions see \citealt{2008A&A...480..735F} for Ext. R CrA and \citealt{2008hsf2.book..757T} for Octans). The five remaining former companion candidates to \rxjs{} are discussed in detail in the following paragraphs. For all of them, the Octans association is the proposed host association of the SN. This is mainly due to the large size of the association.\footnote{Although \citet{2008hsf2.book..757T} question whether the Octans association is expanding, it could well be that it was smaller in the past, i.e. at the time of the SN. However, we neglect the expansion of associations here to not miss candidate stars. The chance of finding former companion candidates is naturally higher for large associations, however we regard each association with a runaway star individually. Every companion star is discussed in detail and further observation of the runaway star might be necessary to confirm or reject a particular SN scenario.} It also still could be that more distant runaway stars that could confirm a SN scenario for \rxjs{} in Sco OB4 are not present in the \citet{2011MNRAS.410..190T} catalogue for runaway star candidates as it is based on Hipparcos data (complete to $\approx\unit[0{.}5]{kpc}$). We stress that the non-identification of a possible former companion for \rxjs{} for a SN in Sco OB4 could be caused by the lack of more distant runaway stars in the runaway star catalogue or by the possibility that the progenitor star was a single star.\\

\paragraph{HIP 57787}
is a metal-poor G5 giant star with a metallicity of [Fe/H]$=-2{.}10\pm0{.}16$ \citep{1994AJ....107.1577A}. Due to the low metallicity, the runaway nature of this star is probably not due to a SN event in a close binary system.

\paragraph{HIP 80448}
is a G2V pre-main sequence binary with a companion that is itself a single-lined spectroscopic binary \citep{2003A&A...397..987C} with an age of $\unit[17{.}2\pm2{.}9]{Myr}$ \citep{2011MNRAS.410..190T}. Its age is consistent with the age of the Octans association ($\unit[20]{Myr}$, \citealt{2008hsf2.book..757T}) which could have hosted the SN.

\paragraph{HIP 81696}
is an O6.5V double star. An encounter of the system with \rxjs{} could have occurred inside the Octans association if the present distance of the runaway star candidate was $\unit[36^{+98}_{-14}]{pc}$. However, distance estimates from Ca-II column densities indicate a distance of the system of $\approx\unit[1]{kpc}$. Also, the Hipparcos parallax of $\unit[17{.}95\pm16{.}68]{mas}$ \citep{2007AA...474..653V} that we used in our simulations is very uncertain. Due to the large distance of HIP 81696, we may exclude it as a companion candidate to \rxjs{}.

\paragraph{HIP 82977}
is an A0 \citep{2006A&A...446..785M} eclipsing binary of Algol type and lies below the theoretical zero-age main sequence (ZAMS) of the models used by \citet{2011MNRAS.410..190T} to determine stellar ages, it was shifted towards the ZAMS and treated as ZAMS star. Therefore, it's age is very uncertain. Since A0 stars spend a few hundred Myrs on the main sequence, it is unlikely that HIP 82997 is only a few tens of Myrs old. Hence, it is probably not the former companion of \rxjs{}.

\paragraph{HIP 86228}
is a variable star with spectral type F1II \citep[see][]{2007AJ....134.1089S}. \citet{1972A&A....19..167O} give a helium abundance for HIP 86228 of $N_{He}/N_H\approx0{.}1$ that is comparable to the mean helium abundance of B type stars \citep[for a review see also][]{2010KPCB...26..169L}. However, they used a spectral type of B1III, hence a much different effective temperature for this star, the given helium abundance might thus be wrong. Re-measuring the helium abundance of HIP 86228 might hence be crucial.\\

\begin{figure}
\centering
(a) HIP 80448
\includegraphics[width=0.4\textwidth, viewport= 40 205 540 600]{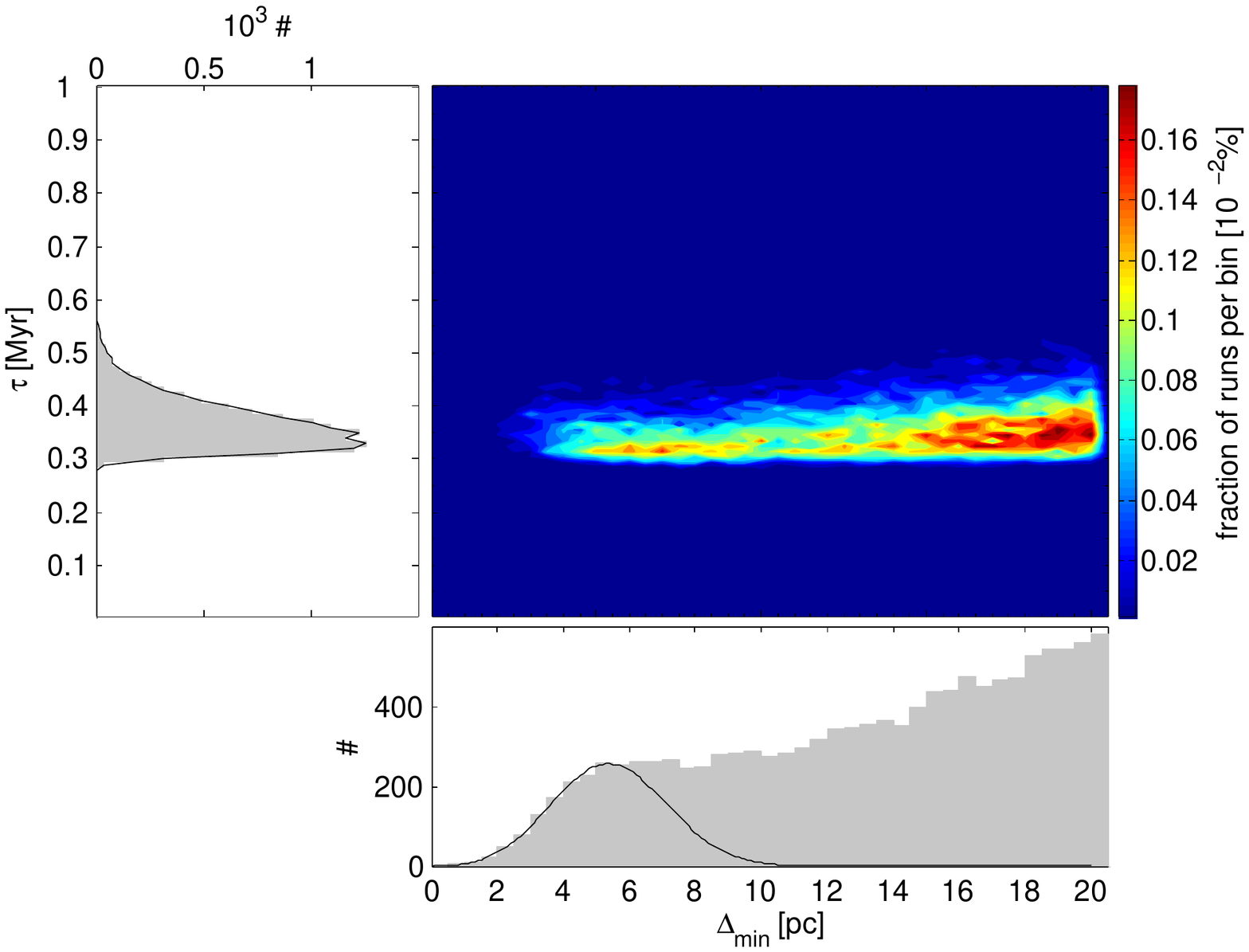}\newline
(b) HIP 86228
\includegraphics[width=0.4\textwidth, viewport= 40 205 540 600]{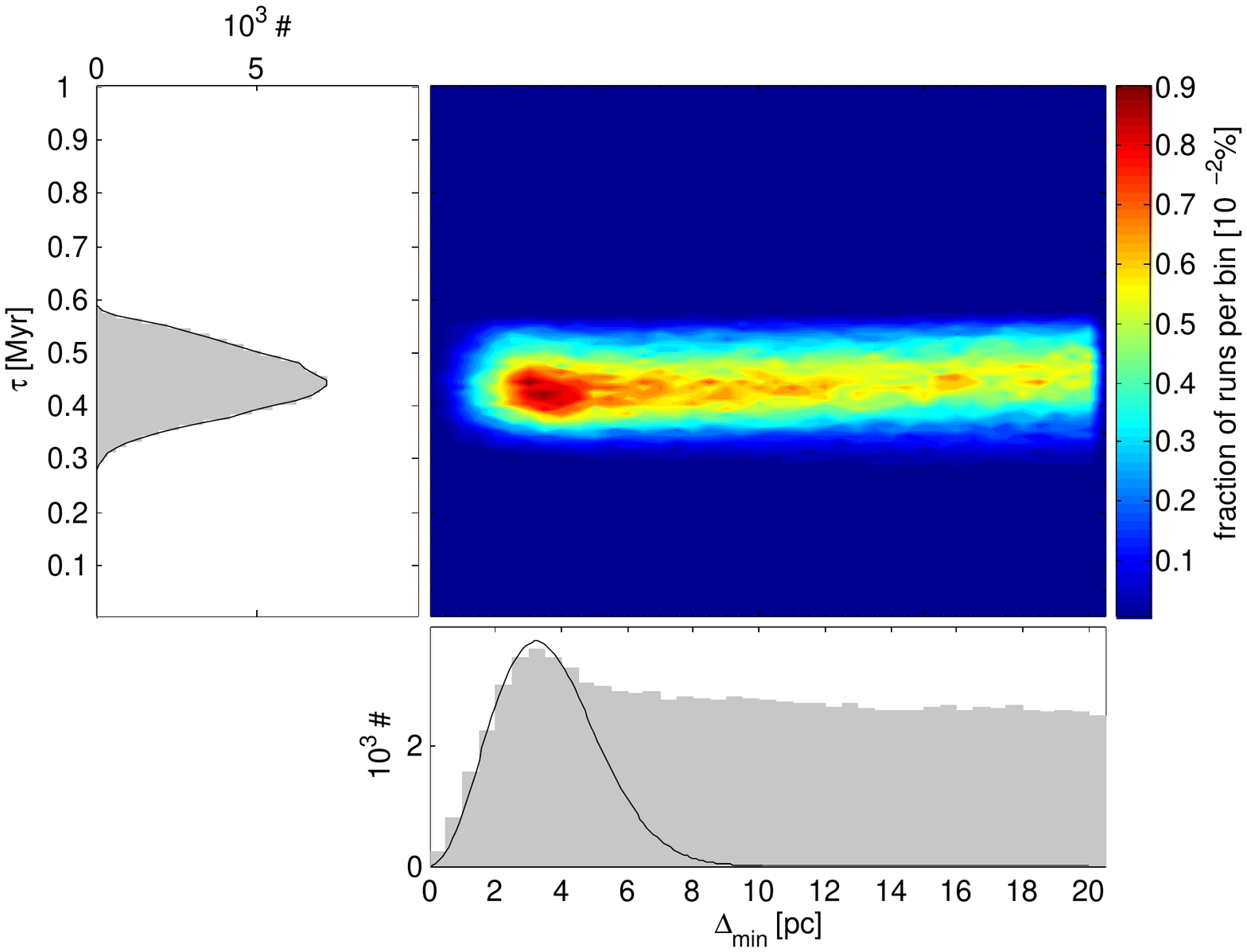}
\caption{Distributions of minimum separations $\Delta_{min}$ and corresponding flight times $\tau$ for encounters between \rxjs{} and the runaway star candidates HIP 80448 (a) and HIP 86228 (b). The solid curves drawn in the $\Delta_{min}$ histograms (bottom panels) represent the theoretically expected distributions (\autorefs{eq:3DGauss_diff} or \ref{eq:3DGaussmunull_diff}), adapted to the first part of each histogram, see \autoref{sec:proc}. For HIP 80448 we find that there could have been a close fly-by ($m=\unit[4{.}7]{pc}$, $s=\unit[1{.}8]{pc}$) at $\unit[0{.}33^{+0{.}05}_{-0{.}02}]{Myr}$ in the past (solid line in the $\tau$ histogram, left panels), however no direct encounter. For HIP 86228 the adapted curve predicts that \rxjs{} and the runaway star could have been at the same place ($m=\unit[0]{pc}$, $s=\unit[2{.}3]{pc}$) at $\unit[0{.}45^{+0{.}06}_{-0{.}06}]{Myr}$ in the past. Both encounters would have occurred inside the Octans association.}\label{fig:runsmitRVplots}
\end{figure}

Considering the individual discussion above, we regard HIP 80448 and HIP 86228 as possible former companion candidates to \rxjs{}. In \autoref{fig:runsmitRVplots} we show the distributions of minimum separations $\Delta_{min}$ and corresponding flight times $\tau$ (the kinematic age of the NS) for those runs where both, the NS and the runaway star were within the boundaries of the Octans association. We adapted the theoretically expected distribution (\autorefs{eq:3DGauss_diff} and \ref{eq:3DGaussmunull_diff}) to the first part of the histogram. For HIP 86228 the curve predicts that the runaway star could have been at the same place at the same time in the past as the NS, whereas for HIP 80448 a fly-by is suggested. Hence HIP 86228 is so far our best former companion candidate for \rxjs{}. 

Since there are also 842 runaway star candidates listed in the \citet{2011MNRAS.410..190T} catalogue without radial velocities available in the literature, we search for possible encounters with \rxjs{} among these stars assuming $v_r=\unit[\pm500]{\kms}$ for the runaway star. For 47 of them it is possible to find a past position as close as $\unit[10]{pc}$ to \rxjs{} after $10^4$ Monte Carlo runs. After further three million runs for those stars, we found 26 candidates with a smallest separation to \rxjs{} less than one parsec and a peculiar space velocity smaller than at least $\unit[180]{\kms}$ (that is $\approx6\sigma$ above the maximum of the distribution of runaway star velocities, \citealt{2011MNRAS.410..190T}). For those stars, we evaluate the reference probability $P_0$\footnote{ At this point, we adopt a velocity distribution for these stars according to \citet{2011MNRAS.410..190T}. Otherwise, the probability of occurrence of values $\Delta_{min}$, $P$, is a priori small and cannot be compared to $P_0$. In the case of NSs, the large variety of possible velocities is already accounted for.} (see above). We find no significant association between a runaway star candidate without known radial velocity and \rxjs{}. We still regard stars for which possible encounters with \rxjs{} are found within one of the potential birth associations of the NS (\autoref{tab:assoc1605_2}) (for them, it is expected to find encounters less significant, as explained earlier), as former companion candidates. For them, we evaluate whether \autoref{eq:3DGaussmunull_diff} could fit the first part of the histogram of separations $\Delta_{min}$ between \rxjs{} and the runaway star (for those Monte Carlo runs yielding small $\Delta_{min}$ values within the potential parent association, see also \autoref{fig:runsmitRVplots} for examples of such distributions). In \autoref{tab:candsohneRV}, we give the parameters of the adapted curves ($m$, $s$) and the corresponding flight times $\tau$. For three runaway stars we find that the NS and runaway star could have been at the same place and time within the Octans association. 

HIP 83003 is listed as an O type Cepheid variable (Simbad database\footnote{http://simbad.u-strasbg.fr/simbad, operated at CDS}). However, its high $V$ band magnitude $V=\unit[10{.}62]{mag}$ suggests that the star might be much more distant than its parallactic distance of $\unit[163^{+68}_{-37}]{pc}$ (infers absolute magnitude of $M_V = \unit[4{.}6]{mag}$; for an O type star, $M_V\approx0$ to $\unit[-5]{mag}$, hence its distance is $\approx\unit[1-13]{kpc}$).

HIP 98500 is an F2 main sequence star. As it lies just below the theoretical model ZAMS, hence was treated as ZAMS star by \citet{2011MNRAS.410..190T}. Taking into account the uncertainty in distance, we estimate its age to be roughly $\unit[600^{+500}_{-400}]{Myr}$ (with ages possible as small as $\approx\unit[50]{Myr}$) from its position in the HRD. Hence, it is probably too old to be the possible former companion of \rxjs{}. Furthermore, HIP 98500 is a metal-weak star \citep{Bidelman}, suggesting that, if it is a true young runaway star, it was ejected due to dynamical interactions in a dense cluster rather than a SN in a binary system. 

HIP 89394 was classified as F0II star by \citet{1975mcts.book.....H}, and revised by \citet{1979A&AS...37..367O} to be Am, i.e. it is metal-rich. We suggest further investigation of that star and treat it as a possible former companion candidate. 

\begin{table}
\centering
\caption{Properties of potential encounters inside Octans between \rxjs{} and our former companion candidates.\normalsize}\label{tab:candsohneRV}
\begin{tabular}{c d{2.1} d{2.1} >{$}c<{$} }
\hline
HIP		&	\multicolumn{1}{c}{$m$ [pc]}		&	\multicolumn{1}{c}{$s$ [pc]}	&	\tau\ $[Myr]$								\\\hline
10137	&	77.0	&	24.9		&	0{.}46^{+0{.}05}_{-0{.}07}	\\
83003	&	0			&	8.9			&	0{.}40^{+0{.}06}_{-0{.}06}	\\
89394	&	0			&	2.3			&	0{.}47^{+0{.}03}_{-0{.}07}	\\
92218	&	5.3		&	1.3			&	0{.}44^{+0{.}04}_{-0{.}06}	\\
95800	&	0			&	2.0			&	0{.}43^{+0{.}05}_{-0{.}05}	\\
\hline
\multicolumn{4}{p{0.4\textwidth}}{Col. 1: Hipparcos identifier\newline
															 Cols. 2,3: parameters of \autorefs{eq:3DGauss_diff} or \ref{eq:3DGaussmunull_diff} adapted to the first part of the $\Delta_{min}$ distributions\newline
															 Col. 4: flight time of the stars (kinematic age of \rxjs{})}
\end{tabular}
\end{table}

\section{Summary and conclusions}\label{sec:summary}

Many NSs and runaway stars apparently come from the same regions on the sky, i.e. regions where massive star associations and clusters are present (\autoref{fig:gammaplot}). To identify birth places of NSs, we attempt to find NS-runaway star pairs that could have been former companions that were disrupted during the SN of the primary. 

Here, we presented our results on the origin of the isolated NS \rxjs{}. We suggest that it was born in the Octans association $\approx\unit[0{.}45]{Myr}$ ago. Such a scenario is supported by the identification of two former companion candidates that are now runaway stars, namely HIP 86228 and HIP 89394. For the latter, no radial velocity has yet been published. 

For both cases, i.e. either HIP 86228 or HIP 89394 being the former companion of \rxjs{}, the predicted current parameters of \rxjs{} and the predicted SN position (coordinates as seen from Earth at present) and time for this case are summarized in \autoref{tab:predpar} (see \citealt{2010MNRAS.402.2369T} for their derivation). 
\begin{table}
\centering
\caption{Predicted current parameters of \rxjs{} and SN position and time.}\label{tab:predpar}
{\footnotesize
\begin{tabular}{l >{\hspace{-5em}}c >{\hspace{-5em}}c}
\hline
former comp. cand.	&	HIP 86228	&	 HIP 89394\\
$v_r$ [$\kms$]             & $588^{+152}_{-111}$	& $626^{+209}_{-56}$	\\
current NS dist. [pc]              & $303^{+42}_{-25}$	& $303^{+54}_{-40}$\\
$\mu_{\alpha}^*$ [mas yr$^{-1}$] & $-43{.}7\pm1{.}6$ & $-43{.}7\pm1{.}7$	\\
$\mu_\delta$ [mas yr$^{-1}$]     & $148{.}4\pm2{.}6$ & $148{.}6\pm2{.}6$	\\
$v_{sp}$ [$\kms$]					 & $587^{+186}_{-75}$			& $657^{+183}_{-84}$			\\
dist. of SN to the Sun [pc] & $92^{+15}_{-11}$	& $99^{+16}_{-20}$\\
Gal. long. $l$ [deg]    & $342{.}9^{+1{.}3}_{-0{.}8}$ & $335{.}6^{+2{.}7}_{-1{.}0}$	\\
Gal. lat. $b$ [deg]     & $-7{.}3^{+1{.}3}_{-0{.}8}$	& $-16{.}7^{+2{.}7}_{-1{.}6}$	\\
time in the past [Myr]   & $0{.}45\pm0{.}06$	& $0{.}43\pm0{.}05$	\\\hline
\multicolumn{2}{p{0.4\textwidth}}{Rows 1-5: NS parameters, Rows 6-9: predicted SN position and time. HIP 89394 would have a radial velocity of $v_{r,HIP}=\unit[117^{+99}_{-34}]{\kms}$}
\end{tabular}
}
\end{table}

As the predicted SNe would have been very recent ($\approx\unit[0{.}45]{Myr}$ ago) and nearby ($\approx\unit[100]{pc}$), we expect to find $\gamma$ ray emission from $^{26}$Al decay \citep[e.g.][]{2010A&A...522A..51D}. In the case of HIP 86228, at the predicted position the emission is dominated by emission from the galactic inner region. Also in the case of HIP 89394, the predicted SN position is not far from the galactic plane; however, in the COMPTEL $\unit[1{.}8]{MeV}$ map \citep[e.g.][]{2010A&A...522A..51D} there is a feature centred at $\left(l, b\right)=\left(334^\circ\hspace{-0.8ex}.5, -16^\circ\hspace{-0.8ex}.6\right)$. According to SN remnant expansion theory (Sedov-Taylor expansion, Snowplough expansion, \citealt{2003pafd.book.....C}), a SN remnant expanding into the interstellar medium with a volume density of $n=\unit[1]{cm^{-3}}$ would have a size of $\approx\unit[46]{pc}$ after $\unit[0{.}43]{Myr}$. At a distance of $\unit[100]{pc}$, this corresponds to an angular size of the SN remnant of $\approx25^\circ$. The size of the feature is close to the COMPTEL resolution ($\unit[1]{deg^2}$). If we treat it as a point source, we obtain a total flux of $\unit[2{.}5\cdot10^{-5}]{photons\,cm^{-1}\,s^{-1}}$. With a half-life of $^{26}$Al of $\unit[0{.}72]{Myr}$ \citep[e.g.][]{1958ZNatA..13..847R,1984PhRvC..30..385T}, this yields an ejected mass of $^{26}$Al in the SN of $\unit[2{.}2\cdot10^{-5}]{M_\odot}$. Compared to theoretical $^{26}$Al yields by \citet{1995ApJS..101..181W} and \citet{2005NuPhA.758...11L}, this corresponds to a mass of the progenitor star of $\approx\unit[11]{M_\odot}$.

\begin{figure}
\centering
\includegraphics[width=0.4\textwidth, viewport= 40 205 540 600]{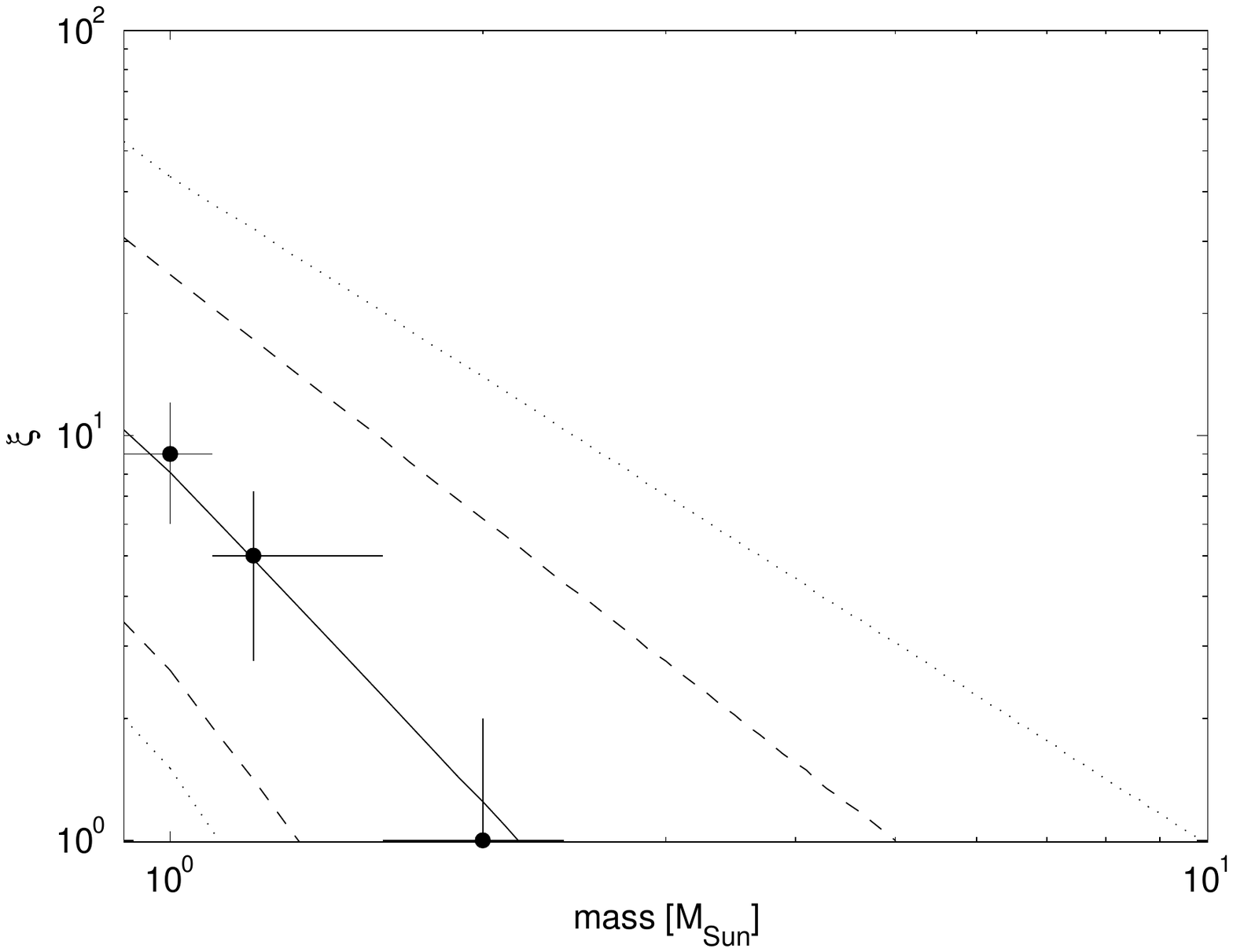}
\caption{Present mass function of the Octans association (black circles) in comparison with the initial mass function $\xi$ (IMF, solid line) as given by \citet{2005ASSL..327..175K}, their equation 2. Dashed and dotted lines represent the $1\sigma$ and $1{.}5\sigma$ IMF boundaries. Within $1{.}5\sigma$ we expect up to one star with $\unit[10]{M_\odot}$; the probability for higher mass stars is low.}\label{fig:Oct_massfun}
\end{figure}

The progenitor mass can also be derived from the life-time of the progenitor star which can be estimated as the difference between the age of the parent association and the age of the NS. With the Octans association being the parent association of the NS and runaway star, we estimate a progenitor mass of $\unit[10-11]{M_\odot}$ for both cases. This is also in excellent agreement with the estimates from $^{26}$Al for the case of HIP 89394. A progenitor mass of $\unit[10-11]{M_\odot}$ corresponds to spectral type B1 on the main sequence. All high probability Octans members in the list by \citet{2008hsf2.book..757T} are F to K type stars, hence very low mass stars. From the Octans present mass function it might be possible within $1{.}5\sigma$ that there was one $\unit[10]{M_\odot}$ star in Octans (\autoref{fig:Oct_massfun}).\\
Our first scenario involving HIP 86228 as former companion candidate might be supported by the existence of the old (few $10^4$ to $\unit[10^5]{yr}$) nearby ($\lesssim\unit[200]{pc}$) SN remnant RCW 114 \citep{1984MNRAS.210..693B,2010ApJ...709..823K} at $\left(l, b\right)\approx\left(334^\circ, -6^\circ\right)$ (\citealt{2009BASI...37...45G}; A. Poghosyan, priv. comm.).\\
We plan to take higher resolution spectra of HIP 86228 and HIP 89394 to measure their (He) abundances, radial velocities and rotational velocities.\\
We note that more distant runaway stars that could support a SN in Sco OB4 might have been not found because of the lack of more distant runaway stars in the runaway star catalogue which will be updated and extended to more distant stars in the future.\\

It has been suggested by \citet{1996ApJ...470.1227E} that a nearby SN should deposit radioactive isotopes such as $^{60}$Fe on Earth. Since our proposed SN  would have been close to Earth ($\approx\unit[100]{pc}$), we may expect to find a signal of $^{60}$Fe. However, there is no significant signal (younger than $\unit[1]{Myr}$) detected in the sediment investigated by \citet{2004PhRvL..93q1103K} and \citet{2008PhRvL.101l1101F}. This may has several reasons: as the ejecta material reached the Earth, it contaminated other parts of the surface than the regions where the sediment investigated by \citet{2004PhRvL..93q1103K} and \citet{2008PhRvL.101l1101F} was found, maybe where the conditions were not suitable to allow such sedimentations. The sediment could be destroyed over the time or it takes more time for the ejecta be up-taken by the ocean and/ or the host material than currently expected. Another possibility is that the amount of $^{60}$Fe that is released in the SN is much less than assumed or the SN is asymmetric. Hence, a non-detection of an $^{60}$Fe Peak does not exclude a nearby SN event at the particular time.

If an identification of the birth place of a NS can be made, the kinematic age gives the best estimate on the true NS age since the spin-down age gives only an upper limit in most cases (see also \citealt{2010MNRAS.402.2369T}). Moreover, for \rxjs{} no spin-down has been measured yet and the age can only be determined kinematically. Our result of $\tau_{kin}\approx\unit[0{.}45]{Myr}$ is in good agreement with cooling models of NSs (see \citealt{2010MNRAS.402.2369T}; for a high magnetic field consisting of polar and toroidal components, cooling curves fit even better, \citealt{2009A&A...496..207P}).

\section*{Acknowledgments} 

We would like to thank the organizers of the Workshop Astronomy with Radioactivities VII held 1-3 March 2011 on Phillip Island, Australia for a very stimulating conference and discussions. 
We are grateful to Roland Diehl for COMPTEL data, suggestions and discussions. We thank Amalya Poghosyan for providing an updated catalogue of SN remnants (not yet public).
We thank the anonymous referee for useful comments.
NT acknowledges Carl-Zeiss-Stiftung for a scholarship. JGS, MMH and RN acknowledge support from DFG in the SFB/TR-7 Gravitational
Wave Astronomy.\\
This work has made use of the Simbad database, http://simbad.u-strasbg.fr/simbad, operated at CDS.


\bibliographystyle{apj}
\bibliography{bib_AwRVII}

\begin{thebibliography}{72}
\expandafter\ifx\csname natexlab\endcsname\relax\def\natexlab#1{#1}\fi

\bibitem[{{Anthony-Twarog} \& {Twarog}(1994)}]{1994AJ....107.1577A}
{Anthony-Twarog}, B.~J., \& {Twarog}, B.~A. 1994, \aj, 107, 1577

\bibitem[{{Arzoumanian} {et~al.}(2002){Arzoumanian}, {Chernoff}, \&
  {Cordes}}]{2002ApJ...568..289A}
{Arzoumanian}, Z., {Chernoff}, D.~F., \& {Cordes}, J.~M. 2002, \apj, 568, 289

\bibitem[{{Asiain} {et~al.}(1999){Asiain}, {Figueras}, \&
  {Torra}}]{1999A&A...350..434A}
{Asiain}, R., {Figueras}, F., \& {Torra}, J. 1999, \aap, 350, 434

\bibitem[{{Bedford} {et~al.}(1984){Bedford}, {Elliott}, {Ramsey}, \&
  {Meaburn}}]{1984MNRAS.210..693B}
{Bedford}, D.~K., {Elliott}, K.~H., {Ramsey}, B., \& {Meaburn}, J. 1984,
  \mnras, 210, 693

\bibitem[{{Belczynski} \& {Taam}(2008)}]{2008ApJ...685..400B}
{Belczynski}, K., \& {Taam}, R.~E. 2008, \apj, 685, 400

\bibitem[{{Bergh{\"o}fer} \& {Breitschwerdt}(2002)}]{2002A&A...390..299B}
{Bergh{\"o}fer}, T.~W., \& {Breitschwerdt}, D. 2002, \aap, 390, 299

\bibitem[{{Bidelman} \& {McConnell}(1978)}]{Bidelman}
{Bidelman}, W.~P., \& {McConnell}, B. 1978, \aj, 78, 687

\bibitem[{{Blaauw}(1961)}]{1961BAN....15..265B}
{Blaauw}, A. 1961, \bain, 15, 265

\bibitem[{{Bobylev}(2008)}]{2008AstL...34..686B}
{Bobylev}, V.~V. 2008, Astronomy Letters, 34, 686

\bibitem[{{Bobylev} \& {Bajkova}(2009)}]{2009AstL...35..396B}
{Bobylev}, V.~V., \& {Bajkova}, A.~T. 2009, Astronomy Letters, 35, 396

\bibitem[{{Burrows} \& {Hayes}(1996)}]{1996PhRvL..76..352B}
{Burrows}, A., \& {Hayes}, J. 1996, Physical Review Letters, 76, 352

\bibitem[{{Chmyreva} {et~al.}(2010){Chmyreva}, {Beskin}, \&
  {Biryukov}}]{2010AstL...36..116C}
{Chmyreva}, E.~G., {Beskin}, G.~M., \& {Biryukov}, A.~V. 2010, Astronomy
  Letters, 36, 116

\bibitem[{{Clarke} \& {Carswell}(2003)}]{2003pafd.book.....C}
{Clarke}, C.~J., \& {Carswell}, R.~F. 2003, {Principles of Astrophysical Fluid
  Dynamics}, ed. {Clarke, C.~J.~\& Carswell, R.~F.}

\bibitem[{{Cordes} \& {Chernoff}(1998)}]{1998ApJ...505..315C}
{Cordes}, J.~M., \& {Chernoff}, D.~F. 1998, \apj, 505, 315

\bibitem[{{Cutispoto} {et~al.}(2003){Cutispoto}, {Tagliaferri}, {de Medeiros},
  {Pastori}, {Pasquini}, \& {Andersen}}]{2003A&A...397..987C}
{Cutispoto}, G., {Tagliaferri}, G., {de Medeiros}, J.~R., {Pastori}, L.,
  {Pasquini}, L., \& {Andersen}, J. 2003, \aap, 397, 987

\bibitem[{{Diehl} {et~al.}(2010){Diehl}, {Lang}, {Martin}, {Ohlendorf},
  {Preibisch}, {Voss}, {Jean}, {Roques}, {von Ballmoos}, \&
  {Wang}}]{2010A&A...522A..51D}
{Diehl}, R., {et~al.} 2010, \aap, 522, A51+

\bibitem[{{Ellis} {et~al.}(1996){Ellis}, {Fields}, \&
  {Schramm}}]{1996ApJ...470.1227E}
{Ellis}, J., {Fields}, B.~D., \& {Schramm}, D.~N. 1996, \apj, 470, 1227

\bibitem[{{Fern{\'a}ndez} {et~al.}(2008){Fern{\'a}ndez}, {Figueras}, \&
  {Torra}}]{2008A&A...480..735F}
{Fern{\'a}ndez}, D., {Figueras}, F., \& {Torra}, J. 2008, \aap, 480, 735 (F08)

\bibitem[{{Fitoussi} {et~al.}(2008){Fitoussi}, {Raisbeck}, {Knie}, \& {et
  al.}}]{2008PhRvL.101l1101F}
{Fitoussi}, C., {Raisbeck}, G.~M., {Knie}, K., \& {et al.} 2008, Physical
  Review Letters, 101, 121101

\bibitem[{{Green}(2009)}]{2009BASI...37...45G}
{Green}, D.~A. 2009, Bulletin of the Astronomical Society of India, 37, 45

\bibitem[{{Gusakov} {et~al.}(2005){Gusakov}, {Kaminker}, {Yakovlev}, \&
  {Gnedin}}]{2005MNRAS.363..555G}
{Gusakov}, M.~E., {Kaminker}, A.~D., {Yakovlev}, D.~G., \& {Gnedin}, O.~Y.
  2005, \mnras, 363, 555

\bibitem[{{Haberl}(2007)}]{2007Ap&SS.308..181H}
{Haberl}, F. 2007, \apss, 308, 181

\bibitem[{{Hansen} \& {Phinney}(1997)}]{1997MNRAS.291..569H}
{Hansen}, B.~M.~S., \& {Phinney}, E.~S. 1997, \mnras, 291, 569

\bibitem[{{Heger} {et~al.}(2003){Heger}, {Fryer}, {Woosley}, {Langer}, \&
  {Hartmann}}]{2003ApJ...591..288H}
{Heger}, A., {Fryer}, C.~L., {Woosley}, S.~E., {Langer}, N., \& {Hartmann},
  D.~H. 2003, \apj, 591, 288

\bibitem[{{Hobbs} {et~al.}(2005){Hobbs}, {Lorimer}, {Lyne}, \&
  {Kramer}}]{2005MNRAS.360..974H}
{Hobbs}, G., {Lorimer}, D.~R., {Lyne}, A.~G., \& {Kramer}, M. 2005, \mnras,
  360, 974

\bibitem[{{Hohle} {et~al.}(2010){Hohle}, {Neuh{\"a}user}, \&
  {Schutz}}]{2010AN....331..349H}
{Hohle}, M.~M., {Neuh{\"a}user}, R., \& {Schutz}, B.~F. 2010, Astronomische
  Nachrichten, 331, 349

\bibitem[{{Hoogerwerf} {et~al.}(2001){Hoogerwerf}, {de Bruijne}, \& {de
  Zeeuw}}]{2001A&A...365...49H}
{Hoogerwerf}, R., {de Bruijne}, J.~H.~J., \& {de Zeeuw}, P.~T. 2001, \aap, 365,
  49

\bibitem[{{Houk} \& {Cowley}(1975)}]{1975mcts.book.....H}
{Houk}, N., \& {Cowley}, A.~P. 1975, {University of Michigan Catalogue of
  two-dimensional spectral types for the HD stars. Volume I.}, ed. {Houk, N.~\&
  Cowley, A.~P.}

\bibitem[{{Janka} \& {Mueller}(1996)}]{1996A&A...306..167J}
{Janka}, H.-T., \& {Mueller}, E. 1996, \aap, 306, 167

\bibitem[{{Janka} {et~al.}(2005){Janka}, {Scheck}, {Kifonidis}, {M{\"u}ller},
  \& {Plewa}}]{2005ASPC..332..363J}
{Janka}, H.-T., {Scheck}, L., {Kifonidis}, K., {M{\"u}ller}, E., \& {Plewa}, T.
  2005, in Astronomical Society of the Pacific Conference Series, Vol. 332, The
  Fate of the Most Massive Stars, ed. R.~{Humphreys} \& K.~{Stanek}, 363--+

\bibitem[{{Kaplan} {et~al.}(2011){Kaplan}, {Kamble}, {van Kerkwijk}, \&
  {Ho}}]{2011ApJ...736..117K}
{Kaplan}, D.~L., {Kamble}, A., {van Kerkwijk}, M.~H., \& {Ho}, W.~C.~G. 2011,
  \apj, 736, 117

\bibitem[{{Kaplan} {et~al.}(2003){Kaplan}, {Kulkarni}, \& {van
  Kerkwijk}}]{2003ApJ...588L..33K}
{Kaplan}, D.~L., {Kulkarni}, S.~R., \& {van Kerkwijk}, M.~H. 2003, \apjl, 588,
  L33

\bibitem[{{Kharchenko} {et~al.}(2005){Kharchenko}, {Piskunov}, {R{\"o}ser},
  {Schilbach}, \& {Scholz}}]{2005A&A...438.1163K}
{Kharchenko}, N.~V., {Piskunov}, A.~E., {R{\"o}ser}, S., {Schilbach}, E., \&
  {Scholz}, R.-D. 2005, \aap, 438, 1163

\bibitem[{{Kim} {et~al.}(2010){Kim}, {Min}, {Seon}, {Han}, \&
  {Edelstein}}]{2010ApJ...709..823K}
{Kim}, I.-J., {Min}, K.-W., {Seon}, K.-I., {Han}, W., \& {Edelstein}, J. 2010,
  \apj, 709, 823

\bibitem[{{Kisslinger} {et~al.}(2009){Kisslinger}, {Henley}, \&
  {Johnson}}]{2009arXiv0906.2802K}
{Kisslinger}, L.~S., {Henley}, E.~M., \& {Johnson}, M.~B. 2009, ArXiv e-prints

\bibitem[{{Knie} {et~al.}(2004){Knie}, {Korschinek}, {Faestermann}, {Dorfi},
  {Rugel}, \& {Wallner}}]{2004PhRvL..93q1103K}
{Knie}, K., {Korschinek}, G., {Faestermann}, T., {Dorfi}, E.~A., {Rugel}, G.,
  \& {Wallner}, A. 2004, Physical Review Letters, 93, 171103

\bibitem[{{Kodama}(1997)}]{1997PhDT........31K}
{Kodama}, T. 1997, PhD thesis, Institute of Astronomy, Univ.~Tokyo

\bibitem[{{Kroupa} \& {Weidner}(2005)}]{2005ASSL..327..175K}
{Kroupa}, P., \& {Weidner}, C. 2005, in Astrophysics and Space Science Library,
  Vol. 327, The Initial Mass Function 50 Years Later, ed. E.~{Corbelli},
  F.~{Palla}, \& H.~{Zinnecker}, 175--+

\bibitem[{{Limongi} \& {Chieffi}(2005)}]{2005NuPhA.758...11L}
{Limongi}, M., \& {Chieffi}, A. 2005, Nuclear Physics A, 758, 11

\bibitem[{{Lorimer} {et~al.}(1997){Lorimer}, {Bailes}, \&
  {Harrison}}]{1997MNRAS.289..592L}
{Lorimer}, D.~R., {Bailes}, M., \& {Harrison}, P.~A. 1997, \mnras, 289, 592

\bibitem[{{Lyne} \& {Lorimer}(1994)}]{1994Natur.369..127L}
{Lyne}, A.~G., \& {Lorimer}, D.~R. 1994, \nat, 369, 127

\bibitem[{{Lyubimkov}(2010)}]{2010KPCB...26..169L}
{Lyubimkov}, L.~S. 2010, Kinematics and Physics of Celestial Bodies, 26, 169

\bibitem[{{Maeder} \& {Meynet}(1989)}]{1989A&A...210..155M}
{Maeder}, A., \& {Meynet}, G. 1989, \aap, 210, 155

\bibitem[{{Malkov} {et~al.}(2006){Malkov}, {Oblak}, {Snegireva}, \&
  {Torra}}]{2006A&A...446..785M}
{Malkov}, O.~Y., {Oblak}, E., {Snegireva}, E.~A., \& {Torra}, J. 2006, \aap,
  446, 785

\bibitem[{{Manchester} {et~al.}(2005){Manchester}, {Hobbs}, {Teoh}, \&
  {Hobbs}}]{2005AJ....129.1993M}
{Manchester}, R.~N., {Hobbs}, G.~B., {Teoh}, A., \& {Hobbs}, M. 2005, \aj, 129,
  1993

\bibitem[{{Mel'Nik} \& {Dambis}(2009)}]{2009MNRAS.400..518M}
{Mel'Nik}, A.~M., \& {Dambis}, A.~K. 2009, \mnras, 400, 518

\bibitem[{{Mereghetti}(2011)}]{2011heep.conf..345M}
{Mereghetti}, S. 2011, in High-Energy Emission from Pulsars and their Systems,
  ed. {D.~F.~Torres \& N.~Rea}, 345--+

\bibitem[{{Motch} {et~al.}(1999){Motch}, {Haberl}, {Zickgraf}, {Hasinger}, \&
  {Schwope}}]{1999A&A...351..177M}
{Motch}, C., {Haberl}, F., {Zickgraf}, F.-J., {Hasinger}, G., \& {Schwope},
  A.~D. 1999, \aap, 351, 177

\bibitem[{{Motch} {et~al.}(2007){Motch}, {Pires}, {Haberl}, \&
  {Schwope}}]{2007Ap&SS.308..217M}
{Motch}, C., {Pires}, A.~M., {Haberl}, F., \& {Schwope}, A. 2007, \apss, 308,
  217

\bibitem[{{Neuh{\"a}user} {et~al.}(2000){Neuh{\"a}user}, {Walter}, {Covino},
  {Alcal{\'a}}, {Wolk}, {Frink}, {Guillout}, {Sterzik}, \&
  {Comer{\'o}n}}]{2000A&AS..146..323N}
{Neuh{\"a}user}, R., {et~al.} 2000, \aaps, 146, 323

\bibitem[{{Olsen}(1979)}]{1979A&AS...37..367O}
{Olsen}, E.~H. 1979, \aaps, 37, 367

\bibitem[{{O'Mara} \& {Simpson}(1972)}]{1972A&A....19..167O}
{O'Mara}, B.~J., \& {Simpson}, R.~W. 1972, \aap, 19, 167

\bibitem[{{Pons} {et~al.}(2009){Pons}, {Miralles}, \&
  {Geppert}}]{2009A&A...496..207P}
{Pons}, J.~A., {Miralles}, J.~A., \& {Geppert}, U. 2009, \aap, 496, 207

\bibitem[{{Posselt} {et~al.}(2007){Posselt}, {Popov}, {Haberl}, {Tr{\"u}mper},
  {Turolla}, \& {Neuh{\"a}user}}]{2007Ap&SS.308..171P}
{Posselt}, B., {Popov}, S.~B., {Haberl}, F., {Tr{\"u}mper}, J., {Turolla}, R.,
  \& {Neuh{\"a}user}, R. 2007, \apss, 308, 171

\bibitem[{{Rightmire} {et~al.}(1958){Rightmire}, {Kohman}, \&
  {Hinten-Berger}}]{1958ZNatA..13..847R}
{Rightmire}, R.~A., {Kohman}, T.~P., \& {Hinten-Berger}, H. 1958, Zeitschrift
  Naturforschung Teil A, 13, 847

\bibitem[{{Rugel} {et~al.}(2009){Rugel}, {Faestermann}, {Knie}, {Korschinek},
  {Poutivtsev}, {Schumann}, {Kivel}, {G{\"u}nther-Leopold}, {Weinreich}, \&
  {Wohlmuther}}]{2009PhRvL.103g2502R}
{Rugel}, G., {et~al.} 2009, Physical Review Letters, 103, 072502

\bibitem[{{Schmidt}(2011)}]{Janos}
{Schmidt}, J. 2011, Diploma thesis, AIU, Friedrich-Schiller-Universit\"at Jena,
  Germany

\bibitem[{{Sowell} {et~al.}(2007){Sowell}, {Trippe}, {Caballero-Nieves}, \&
  {Houk}}]{2007AJ....134.1089S}
{Sowell}, J.~R., {Trippe}, M., {Caballero-Nieves}, S.~M., \& {Houk}, N. 2007,
  \aj, 134, 1089

\bibitem[{{Tetzlaff}(2009)}]{2009DiplA...Nina}
{Tetzlaff}, N. 2009, Diploma thesis, AIU, Friedrich-Schiller-Universit\"at
  Jena, Germany

\bibitem[{{Tetzlaff}(2012)}]{2012DokA...Nina}
---. 2012, PhD thesis, in preparation, AIU, Friedrich-Schiller-Universit\"at
  Jena, Germany

\bibitem[{{Tetzlaff} {et~al.}(2011{\natexlab{a}}){Tetzlaff}, {Eisenbeiss},
  {Neuh{\"a}user}, \& {Hohle}}]{2011arXiv1107.1673T}
{Tetzlaff}, N., {Eisenbeiss}, T., {Neuh{\"a}user}, R., \& {Hohle}, M.~M.
  2011{\natexlab{a}}, \mnras, 417, 617

\bibitem[{{Tetzlaff} {et~al.}(2009){Tetzlaff}, {Neuh{\"a}user}, \&
  {Hohle}}]{2009MNRAS.400L..99T}
{Tetzlaff}, N., {Neuh{\"a}user}, R., \& {Hohle}, M.~M. 2009, \mnras, 400, L99

\bibitem[{{Tetzlaff} {et~al.}(2011{\natexlab{b}}){Tetzlaff}, {Neuh{\"a}user},
  \& {Hohle}}]{2011MNRAS.410..190T}
---. 2011{\natexlab{b}}, \mnras, 410, 190

\bibitem[{{Tetzlaff} {et~al.}(2010){Tetzlaff}, {Neuh{\"a}user}, {Hohle}, \&
  {Maciejewski}}]{2010MNRAS.402.2369T}
{Tetzlaff}, N., {Neuh{\"a}user}, R., {Hohle}, M.~M., \& {Maciejewski}, G. 2010,
  \mnras, 402, 2369

\bibitem[{{Thomas} {et~al.}(1984){Thomas}, {Rau}, {Skelton}, \&
  {Kavanagh}}]{1984PhRvC..30..385T}
{Thomas}, J.~H., {Rau}, R.~L., {Skelton}, R.~T., \& {Kavanagh}, R.~W. 1984,
  \prc, 30, 385

\bibitem[{{Tinsley}(1980)}]{1980FCPh....5..287T}
{Tinsley}, B.~M. 1980, Fundamentals of Cosmic Physics, 5, 287

\bibitem[{{Torres} {et~al.}(2008){Torres}, {Quast}, {Melo}, \&
  {Sterzik}}]{2008hsf2.book..757T}
{Torres}, C.~A.~O., {Quast}, G.~R., {Melo}, C.~H.~F., \& {Sterzik}, M.~F. 2008,
  {Young Nearby Loose Associations}, ed. {Reipurth, B.}, 757--+

\bibitem[{{van Kerkwijk} {et~al.}(2004){van Kerkwijk}, {Kaplan}, {Durant},
  {Kulkarni}, \& {Paerels}}]{2004ApJ...608..432V}
{van Kerkwijk}, M.~H., {Kaplan}, D.~L., {Durant}, M., {Kulkarni}, S.~R., \&
  {Paerels}, F. 2004, \apj, 608, 432

\bibitem[{{van Leeuwen}(2007)}]{2007AA...474..653V}
{van Leeuwen}, F. 2007, \aap, 474, 653

\bibitem[{{Wang} {et~al.}(2006){Wang}, {Lai}, \& {Han}}]{2006ApJ...639.1007W}
{Wang}, C., {Lai}, D., \& {Han}, J.~L. 2006, \apj, 639, 1007

\bibitem[{{Woosley} \& {Weaver}(1995)}]{1995ApJS..101..181W}
{Woosley}, S.~E., \& {Weaver}, T.~A. 1995, \apjs, 101, 181

\bibitem[{{Zane} {et~al.}(2006){Zane}, {de Luca}, {Mignani}, \&
  {Turolla}}]{2006A&A...457..619Z}
{Zane}, S., {de Luca}, A., {Mignani}, R.~P., \& {Turolla}, R. 2006, \aap, 457,
  619

\end{thebibliography}


\appendix

\section{On choosing $\Delta_{min}$ limits}\label{appsec:numMCruns}

We construct a case such that the NS (assuming zero radial velocity for \rxjs{}) and a star with typical runaway star parameters occupied the same place $\unit[1]{Myr}$ in the past and test how small $\Delta_{min}$ can be achieved in the Monte Carlo simulation applying typical errors on the kinematic parameters. We investigate the NS and a simulated runaway star with typical runaway star kinematics as well as one with an assumed radial velocity of $\unit[0\pm500]{\kms}$, i.e. our adopted range when $v_r$ is not known. \Autoref{tab:art_input} gives the adopted $U$, $V$ and $W$ values for the runaway stars in each case.\\
\begin{table}
\centering
\caption{Adopted heliocentric velocities for simulated runaway stars for four different cases.}\label{tab:art_input}
\begin{tabular}{l o{2.2} o{2.2} o{2.2}}
\hline
Case		&	\multicolumn{1}{c}{$U$ [$\kms$]}	&	\multicolumn{1}{c}{$V$ [$\kms$]}	&	\multicolumn{1}{c}{$W$ [$\kms$]}	\\\hline
A				& -11+5														&	-10+5														&	-5+5\\
B				&	-11+435													&	-10+100													&	-5+225\\
\hline
\multicolumn{4}{p{0.4\textwidth}}{A -- \rxjs{} and a runaway star at $\unit[425^{+310}_{-125}]{pc}$ with fully known 3D kinematics, B -- \rxjs{} and a runaway star at $\unit[425^{+310}_{-125}]{pc}$ with an adopted radial velocity of $\unit[0\pm500]{\kms}$.}
\end{tabular}
\end{table}
\begin{figure}
\centering
\includegraphics[width=0.45\textwidth, viewport= 40 205 540 600]{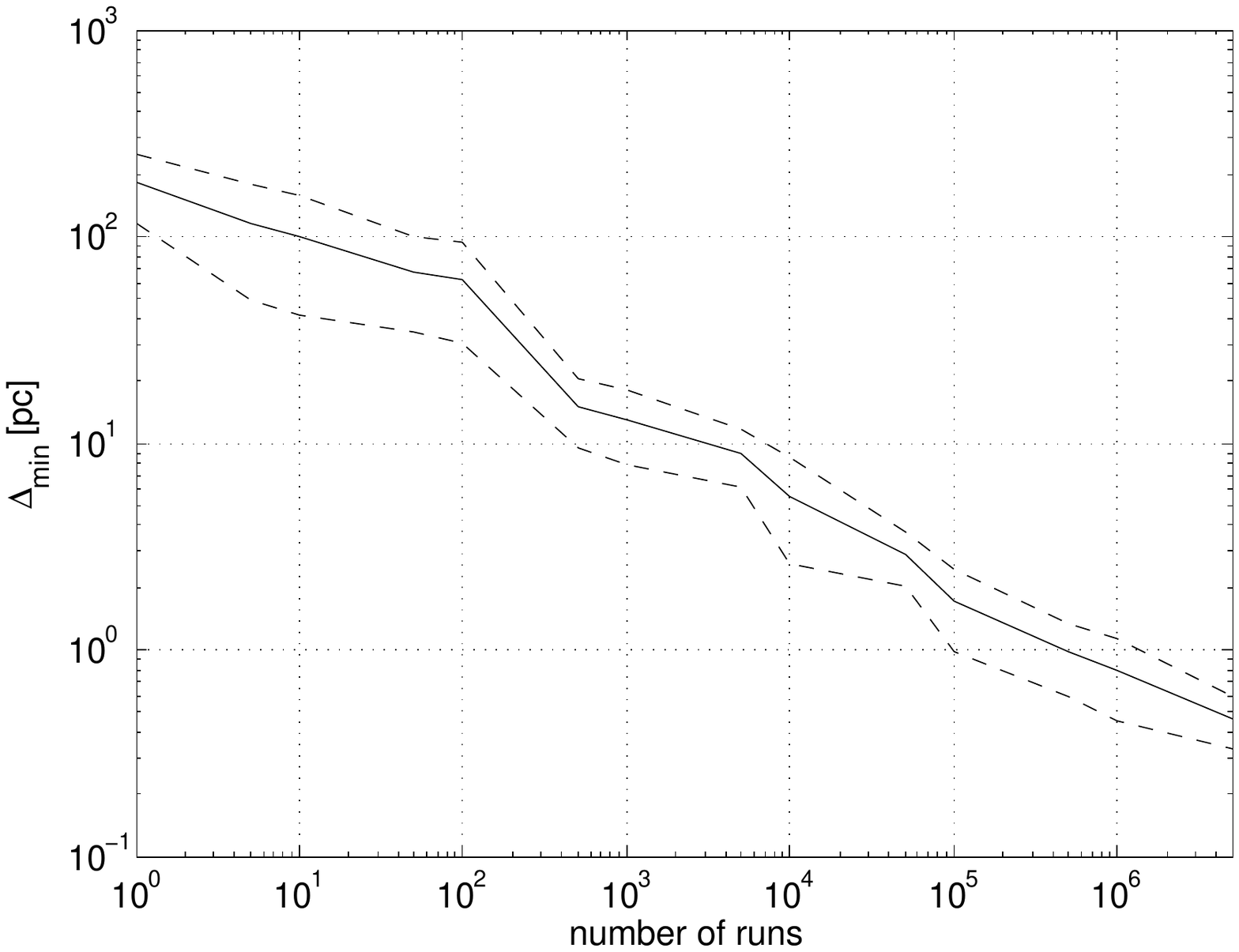}
\includegraphics[width=0.45\textwidth, viewport= 40 205 540 600]{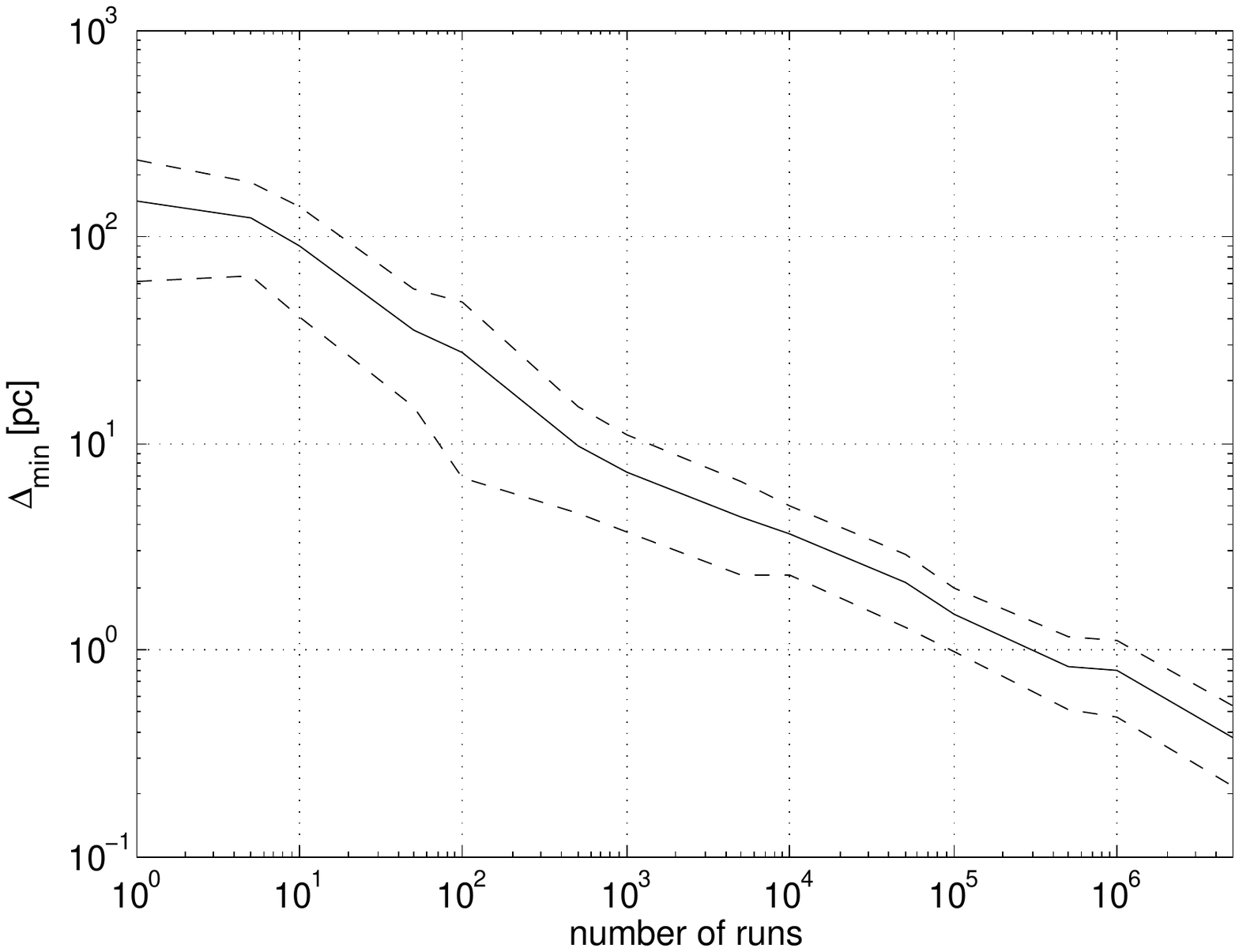}
\caption{Evolution of $\Delta_{min}$ during a Monte Carlo simulation for constructed cases where the NS and the runaway star were at the same place $\unit[1]{Myr}$ in the past. Dashed lines mark the standard deviation. Top panel: Case A -- \rxjs{} and a runaway star at $425^{+310}_{-125}\,\mathrm{pc}$ with fully known 3D kinematics. Bottom panel: Case B -- \rxjs{} and a runaway star at $425^{+310}_{-125}\,\mathrm{pc}$ with an adopted radial velocity of $0\pm500\,\mathrm{\kms}$.}\label{fig:dmin_min}
\end{figure}
Already after $10^3-10^4$ Monte Carlo runs separations below $\unit[10]{pc}$ are reached (\autoref{fig:dmin_min}), below $\unit[1]{pc}$ after $\approx10^6$ runs. To be conservative we carried out many more runs ($10^6-3\times10^6$) to not miss runaway stars near the limits. 

\end{document}